\documentclass[
    ,final            
  ]
  {aipproc}

\usepackage{times,amssymb,mathptmx}

\layoutstyle{6x9}

\begin{document}

\title{Understanding nuclear ``pasta'': current status and future prospects}


\author{Gentaro Watanabe}{
  address={NORDITA, Blegdamsvej 17, DK-2100 Copenhagen \O, Denmark}
  ,altaddress={The Institute of Chemical and Physical Research (RIKEN),
Saitama 351-0198, Japan}
}

\begin{abstract}
In cores of supernovae and crusts of neutron stars,
nuclei can adopt interesting shapes, such as rods or slabs, etc.,
which are referred to as nuclear ``pasta.''
In recent years, we have studied the pasta phases focusing
on their dynamical aspects with a quantum molecular dynamic (QMD) approach.
In this article, we review these works.
We also focus on the treatment of the Coulomb interaction.
\end{abstract}

\maketitle


\section{I.\quad Introduction}

In ordinary matter on the earth, atomic nuclei are roughly spherical.
This fact can be understood in the liquid drop picture of the nucleus 
as being a result of the surface tension of nuclear matter,
which favors a spherical nucleus, being greater than the Coulomb repulsion
between protons, which tends to make the nucleus deform.
However, in supernova cores and neutron stars,
the situation changes completely (see, e.g., Refs.\ \cite{review,chamel}): 
when the density of matter approaches
that of atomic nuclei, i.e., the normal nuclear density $\rho_0$, 
nuclei are closely packed and the effect of the electrostatic energy 
becomes comparable to that of the surface energy.
Consequently, around a density $\rho\simeq\rho_0/2$, 
the energetically favorable
configuration could be rod-like or slab-like nuclei embedded in the gas phase,
or rod-like or spherical bubbles of the gas phase embedded in nuclear matter
\cite{rpw,hashimoto}.
Such phases with exotic shapes of nuclei are referred to as nuclear 
``pasta'' phases.

Properties of the pasta phases in equilibrium states have been investigated 
by many authors so far
\cite{williams,lassaut,lorenz,oyamatsu,sumiyoshi,gentaro}.
These earlier works have confirmed that, for various nuclear models,
the nuclear shape changes in the sequence
sphere, cylinder, slab, cylindrical hole, spherical hole, uniform
with increasing density.
This conclusion holds for the case of non-zero temperatures 
with a constant entropy \cite{lassaut} 
as well as for the zero temperature case.
However, in these works (except for Ref.\ \cite{williams}) 
they took account of only several specific nuclear 
shapes and determined the energetically favorable one 
by calculating the energy for these assumed structures
within a liquid drop model or the Thomas-Fermi approximation.
Thus the phase diagram at subnuclear densities
and the existence of the pasta phases should be examined
without assuming the nuclear shape.
It is also noted that
at typical temperatures of the collapsing cores, several MeV,
effects of thermal fluctuations on the nucleon distribution are significant.
However, these thermal fluctuations cannot be described properly by
mean-field theories such as the Thomas-Fermi and the Hartree-Fock 
approximations.

In the processes of supernova explosion and 
succeeding neutron star formation, the pasta phases can be formed 
in two stages. During the gravitational collapse, matter in the collapsing
core is compressed and the central region starts to solidify into a bcc lattice
because the Coulomb repulsion among nuclei gets stronger.
In the later stage of the collapse, the central density reaches $\rho_0$
and the pasta phases could be formed from the bcc lattice 
due to the compression.
After a bounce of the core takes place, the temperature of the core
increases to $O(10)$ MeV and nuclei melt completely.
As the cooling process of the protoneutron star proceeds later on,
the pasta phases as well as normal spherical nuclei could be produced again
from hot uniform nuclear matter at subnuclear densities.

The previous studies, which are based on a comparison of the energy 
with an assumption of the nuclear shape, cannnot answer
whether or not the pasta phases can be formed dynamically within the time 
scale of the neutron star cooling 
nor whether or not the transitions between them,
which is accompanied by drastic changes of nuclear shape, can be 
realized under nonequilibrium conditions in the collapsing cores.
To solve these problems, molecular dynamic methods
for nucleon many-body systems are suitable.
They treat the motion of the nucleonic degrees of freedom
and can describe thermal fluctuations and many-body correlations
beyond the mean-field level.

Using one of the molecular dynamic methods, the quantum molecular dynamics
(QMD) \cite{aichelin}, we have arrived at the two following major conclusions
\cite{qmd_cold,qmd_hot,qmd_transition}.
(1) The pasta phases are formed from hot uniform nuclear matter 
by cooling it down within a time scale of $\sim O(10^{3}-10^{4})$ fm$/c$.
This supports the idea that the pasta phases exist in neutron star crusts.
(2) The transition from rod-like nuclei to slab-like nuclei and that from
slab-like nuclei to rod-like bubbles can be realized by compression of matter.
This suggests that the pasta phases could be formed in collapsing cores.

Throughout the present article, we set the Boltzmann constant $k_{\rm B}=1$.

\section{II.\quad Method: Quantum Molecular Dynamics}

Among various versions of the molecular dynamic models,
QMD \cite{aichelin} is the most practical and suitable 
for studying the pasta phases.
This method is so efficient that we can describe rod-like and slab-like nuclei
in terms of nucleon degrees of freedom even though they consist of
macroscopic number of nucleons.
In addition, we focus on the pasta structure and the configuration of nuclei
on a macroscopic scale, where the shell effects 
\cite{shell p,shell n,chamel_shell}, which cannot be
described by QMD, may be less important
(in the case of supernova matter, shell effects are always small).
Thus QMD is a good approximation for our present problem.

\subsection{1.\quad Model Hamiltonian}

In our studies, we use a nuclear force
given by a QMD model Hamiltonian with the medium-equation-of-state
parameter set in Ref.\ \cite{maruyama}.
This Hamiltonian contains the momentum-dependent ``Pauli potential'' 
to reproduce the effects of the Pauli principle phenomenologically.
The Pauli potential generates repulsive forces between two identical particles
close together in phase space.
Parameters in the Pauli potential are determined 
by fitting the kinetic energy of the free Fermi gas at zero temperature.

This Hamiltonian reproduces the binding energy of 
symmetric nuclear matter,
16 MeV per nucleon, at the normal nuclear density $\rho_0=0.165$ fm$^{-3}$
and other saturation properties: the incompressibility is set to be
280 MeV and the symmetry energy is 34.6 MeV \cite{maruyama}.
This Hamiltonian also well reproduces the properties of stable nuclei
relevant for the situations we consider:
the binding energy (except for light nuclei from $^{12}$C to $^{20}$Ne) 
\cite{maruyama},
and the rms radius of the ground state of heavy nuclei with $A \gtrsim 100$
\cite{kido}.
It is also confirmed that another QMD Hamiltonian close to this model
provides a good description of nuclear reactions including the low energy
region (several MeV per nucleon) \cite{niita}, which is one of the essential
elements when one studies the dynamical processes as in Sect. III-2.

\subsection{2.\quad Coulomb Interaction and its Screening Effect}

The Coulomb interaction is one of the essential elements for the pasta phases.
One must carefully treat the long-range nature of the Coulomb interaction
since electron screening is negligibly small in this system 
\cite{review,screening,maruyama_screen}.
A key quantity in determining the importance
of electron screening is the ratio of the scale
of the inhomogeneity, typically half the internuclear spacing $R$,
to the Thomas-Fermi screening length $\lambda_{\rm TF}
=[4\pi e^2 (\partial n_e^{(0)}/\partial\mu_e^{(0)})_{n_e^{(0)}}]^{-1/2}
=\sqrt{\pi/4\pi}(k_e^{(0)})^{-1}$,
where $n_e^{(0)}$ and $\mu_e^{(0)}$ are the averaged electron number density
and chemical potential, respectively, and $k_e^{(0)}=\sqrt{3\pi^2 n_e^{(0)}}$
is the electron Fermi wave vector.
We note that, in the density region of the pasta phases, 
$R/\lambda_{\rm TF}<1$ \cite{bbp,review}, which means that it is 
a good approximation to assume that the electron density distribution is
uniform and neglect the screening effect; but one must take account of
the long-range nature of the Coulomb force by, e.g., the Ewald summation 
procedure \cite{qmd_cold}.

It is fortunate that the screening effect is small; otherwise one has to
solve equations of motion for the electronic degrees of freedom together
with that for the nucleonic ones because the coupling between these two
components cannot be neglected.
Thus introducing a screening length shorter than the actual value for
calculating the Coulomb force between protons keeping the density
of electrons uniform is inconsistent.
Let us restrict ourselves to the equilibrium state and
discuss this point more quantitatively 
using a liquid drop model in the Wigner-Seitz approximation.
The energy density of a relativistic degenerate electron gas 
with an inhomogeneous density distribution 
$n_e({\bf r})=n_e^{(0)}+\delta n_e({\bf r})$ is given by
$E_e[n_e({\bf r})]\equiv E_e^{(0)}(n_e^{(0)})+\delta E_e
\simeq E_e^{(0)}(n_e^{(0)})[1+(2/9V_{\rm c})
\int_{\rm cell}d^3r\ (\delta n_e({\bf r})/n_e^{(0)})^2]$, with
$E_e^{(0)}(n_e^{(0)})=(3/4) \hbar ck_e^{(0)}n_e^{(0)}$,
where $V_{\rm c}$ is the volume of the Wigner-Seitz cell.
If the screening effect is significant, the Coulomb energy among protons
decreases, but on the other hand,
the $(\delta n_e({\bf r})/n_e^{(0)})^2$ term cannot be neglected.
Calculations assuming a small $\lambda_{\rm TF}$ and uniform $n_e$
artificially underestimate the Coulomb energy among protons but 
discard the significant contribution of $\delta E_e$.
Now we estimate $\delta E_e$ for an unrealistically small value 
of $\lambda_{\rm TF}$. We take the phase with spherical nuclei
in the density region where the pasta phases start to appear, i.e.,
the volume fraction of nuclei is $u=(r_{\rm N}/r_{\rm c})^3\simeq 1/8$
according to the condition of the fission instability of 
spherical nuclei \cite{review}. Here $r_{\rm N}$ and $r_{\rm c}$ are
the radii of nuclei and the Wigner-Seitz cell, respectively.
Taking supernova matter for simplicity, in which the density of dripped
neutrons is negligible, with the nucleon density in nuclei $n=0.1$ fm$^{-3}$,
proton fraction in nuclei $x_{\rm N}=0.3$, $r_{\rm c}=15$ fm, and
$r_{\rm N}=15/2$ fm, and using expressions in Ref.\ \cite{screening},
$\delta E_e$ per nucleon, $\delta E_e/n_{\rm N}$, is estimated as
$\delta E_e/n_{\rm N}\simeq0.23$ MeV 
\footnote{This calculation is based on the linearized Thomas-Fermi 
approximation and thus the true value of $\delta E_e/n_{\rm N}$ would be larger
than this value.}
for $\lambda_{\rm TF}=10$ fm (this value is adopted in Ref.\ \cite{horowitz}).
Note that this is much larger than the energy difference between
different pasta phases in the same density region in supernova matter, 
$O(10)$ keV per nucleon.

\section{III.\quad Simulations and Results}

In the present section, let us review our previous works 
\cite{qmd_cold,qmd_hot,qmd_transition}.
In our QMD simulations,
we treated a system which consists of neutrons, protons, and electrons
in a cubic box with periodic boundary conditions.
The system is not magnetically polarized,
i.e., it contains equal numbers of protons (and neutrons) with spin up and
spin down.
The relativistic degenerate electrons which ensure charge neutrality
are regarded as a uniform background \cite{review,screening,maruyama_screen}.
The Coulomb interaction is calculated by the Ewald method 
taking account of the Gaussian charge distribution of the proton wave packets.

\subsection{1.\quad Realization of the Pasta Phases
and Equilibrium Phase Diagrams\label{subsect_realization}}

In Refs.\ \cite{qmd_cold,qmd_hot}, we showed that the pasta phases are produced
from hot uniform nuclear matter by cooling it down and we studied 
phase diagrams at zero and non-zero temperatures.
In these works, we first prepared a uniform hot nucleon gas
at a temperature $T \sim 20$ MeV. 
We then cooled it down slowly until the temperature got $\sim 0.1$ MeV
or less for $O(10^{3}-10^{4})$ fm$/c$,
keeping the nucleon number density constant.
Note that this cooling time scale is much larger than the time scale 
$\tau_{\rm relax}$ for relaxation of the system, which is estimated as
$\tau_{\rm relax}\sim$ (length scale of the inhomogeneity)$/$(sound velocity)
$\sim 10\ {\rm fm}/0.1\ c= 100$ fm$/c$.
In the cooling process, we mainly used the frictional relaxation method, 
which is given by the QMD equations of motion plus small friction terms:
$\dot{\bf R}_i=\partial {\cal H}/\partial {\bf P}_i-\xi_R\partial{\cal H}/\partial {\bf R}_i ,$  
$\dot{\bf P}_i=-\partial {\cal H}/\partial {\bf R}_i-\xi_P\partial{\cal H}/\partial {\bf P}_i ,$ where $\cal H$ is the QMD Hamiltonian,
${\bf R}_i$ and ${\bf P}_i$ are the position
and momentum of nucleon $i$, respectively, and $\xi_R, \xi_P>0$ are
the frictional coefficients.

\begin{figure}[t]
\resizebox{11.5cm}{!}{\includegraphics{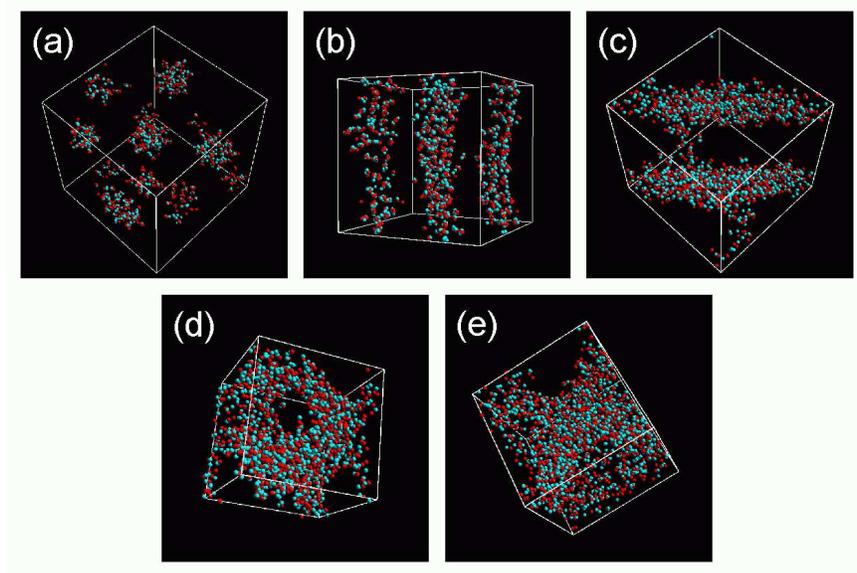}}
\caption{\label{fig pasta sym}
  Nucleon distributions of the pasta phases in cold matter at $x=0.5$;
  (a) sphere phase, $0.1 \rho_{0}$; 
  (b) cylinder phase, $0.225 \rho_{0}$; 
  (c) slab phase, $0.4 \rho_{0}$; 
  (d) cylindrical hole phase, $0.5 \rho_{0}$; 
  (e) spherical hole phase, $0.6 \rho_{0}$. 
  The red (darker) particles represent protons and the green (brighter) 
  ones neutrons.
  Taken from Ref.\ \cite{qmd_cold}.
  }
\end{figure}

The resulting nucleon distributions of cold matter 
at $x=0.5$ are shown in Fig.\ \ref{fig pasta sym}.
We see from these figures that
the phases with rod-like and slab-like nuclei,
cylindrical and spherical bubbles,
in addition to the phase with spherical nuclei are reproduced.
The above simulations show that the pasta phases 
can be formed dynamically from hot uniform matter
within a time scale $\sim O(10^{3}-10^{4})$ fm$/c$.

\begin{figure}[htbp]
\resizebox{13cm}{!}
{\includegraphics{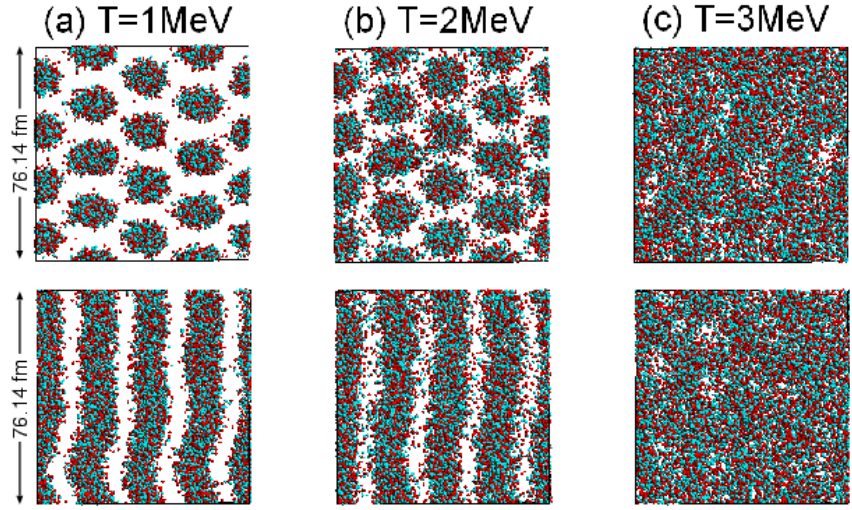}}
\caption{\label{snap 0.225rho x0.5 16000}
  Nucleon distributions for $x=0.5$, $\rho=0.225\rho_{0}$
  at temperatures of 1, 2 and 3 MeV.
  The total number of nucleons $N=16384$ 
  and the box size $L_{\rm box}=76.14$ fm.
  The upper panels show top views along the axis of 
  the cylindrical nuclei at $T=0$, and the lower ones show side views.
  Protons are represented by the red (darker) particles, and
  neutrons by the green (lighter) ones.
  Taken from Ref.\ \cite{qmd_hot}.
  }
\end{figure}

In Fig.\ \ref{snap 0.225rho x0.5 16000}, we show snapshots of the nucleon
distributions for $\rho=0.225\rho_0$ at $T=1, 2$ and 3 MeV.
This density corresponds to the phase with rod-like nuclei at $T=0$.
We have observed the following qualitative features:
at $T\simeq 1.5-2$ MeV
the number of evaporated nucleons becomes significant; 
at $T\gtrsim 3$ MeV, 
nuclei almost melt and the nuclear surface is hard to identify.

\begin{figure}[htbp]
\rotatebox{270}{
\resizebox{7.5cm}{!}
{\includegraphics{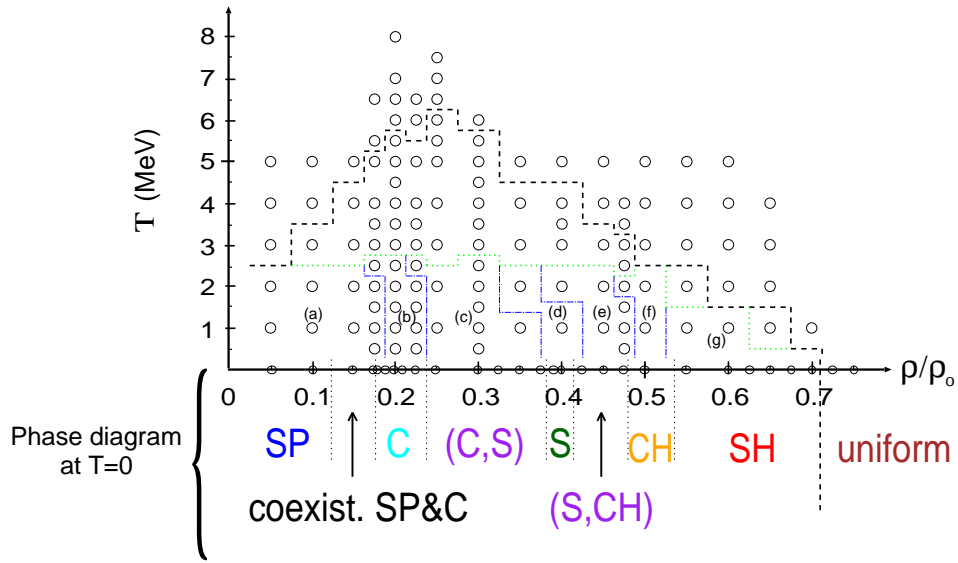}}}
\caption{\label{phase diagram x0.5}
  Phase diagram of matter at $x=0.5$ plotted in the $\rho$ - $T$ plane.
  The dashed and the dotted lines on the diagram
  show the phase separation line and
  the limit below which the nuclear surface can be identified, respectively.
  The dash-dotted lines are the phase boundaries between
  the different nuclear shapes.
  The symbols SP, C, S, CH, SH, U stand for nuclear shapes,
  i.e., sphere, cylinder, slab, cylindrical hole,
  spherical hole and uniform, respectively.
  The parentheses (A,B) denote an intermediate phase between A and B-phases
  with a multiply connected structure. 
  Simulations have been carried out at points denoted by circles.
  Adapted from Ref.\ \cite{qmd_hot}.
  }
\end{figure}

The phase diagram for $x=0.5$ is plotted in 
Fig.\ \ref{phase diagram x0.5}.
The critical temperature of this model is $\gtrsim 6$ MeV.
In the region below the dotted lines at $T \lesssim 3$ MeV,
where we can identify the nuclear surface, we have obtained the pasta phases
in the same sequence as in the earlier works: from lower densities,
spherical nuclei [region (a)], rod-like nuclei [region (b)], 
slab-like nuclei [region (d)],
cylindrical holes [region (f)], and spherical holes [region (g)].
In addition to these pasta phases,
structures with multiply connected nuclear and bubble regions 
(i.e., sponge-like structure) such as
branching rod-like nuclei, perforated slabs and branching bubbles, etc.,
have been obtained in the regions (c) and (e).
Further study using a larger system is necessary to conclude 
the existence of these sponge-like phases 
(see also Refs.\ \cite{review,soft_review,matsuzaki}).

\subsection{2.\quad Structural Transitions between the Pasta Phases
\label{subsect_transition}}

In Ref.\ \cite{qmd_transition}, we demonstrated the second point
mentioned at the beginning of this article.
We performed QMD simulations of the compression of dense matter
and observed the transitions from rod-like nuclei to slab-like ones
and from slab-like nuclei to rod-like bubbles.

The initial conditions of the simulations are samples of the phase with
rod-like nuclei ($\rho=0.225 \rho_0$) and of the phase with slab-like
nuclei ($\rho=0.4 \rho_0$) with 16384 nucleons at 
$x=0.5$ and $T\simeq1$ MeV.
We adiabatically compressed these samples by increasing the density 
to the value corresponding to the next pasta phase
taking $O(10^4)$ fm$/c$.
This time scale is much larger than the typical time scale of the deformation
and structural transition of nuclei (e.g., that of nuclear fission is
$\sim 1000$ fm$/c$). Therefore, the compression in the simulations
is adiabatic with respect to the change of the nuclear structure,
so that the dynamics of the structural transition of nuclei observed
in the simulations is physically meaningful, and is essentially 
independent of the compression rate, etc.

\begin{figure}[t]
\resizebox{12.3cm}{!}
{\includegraphics{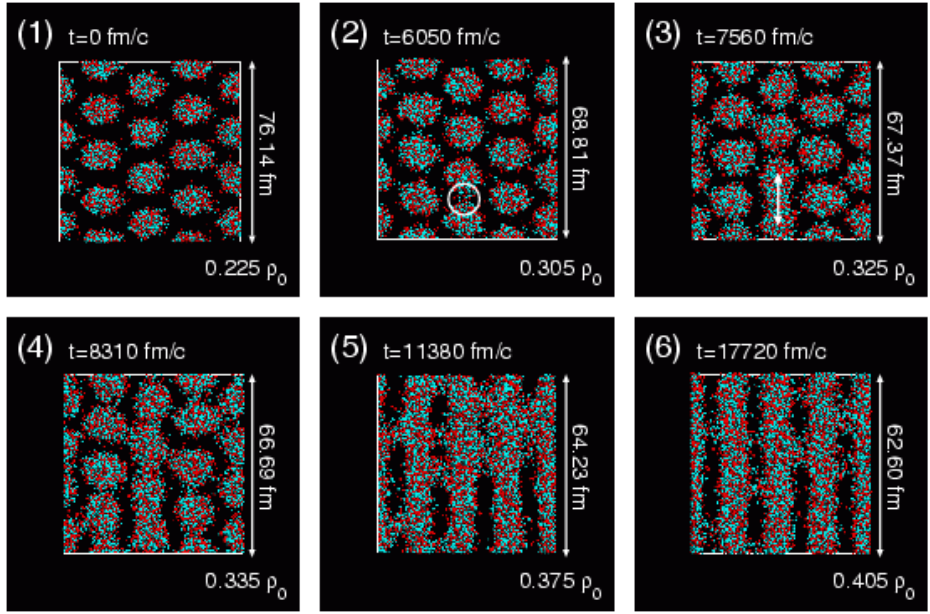}}
\caption{\label{fig rod slab}
Snapshots of the transition process from
the phase with rod-like nuclei to the phase with slab-like nuclei.
The red (darker) particles show protons and the green (lighter) ones neutrons.
The box size is rescaled to be equal in this figure.
Adapted from Ref.\ \cite{qmd_transition}.
}
\end{figure}

\begin{figure}[t]
\resizebox{12.3cm}{!}
{\includegraphics{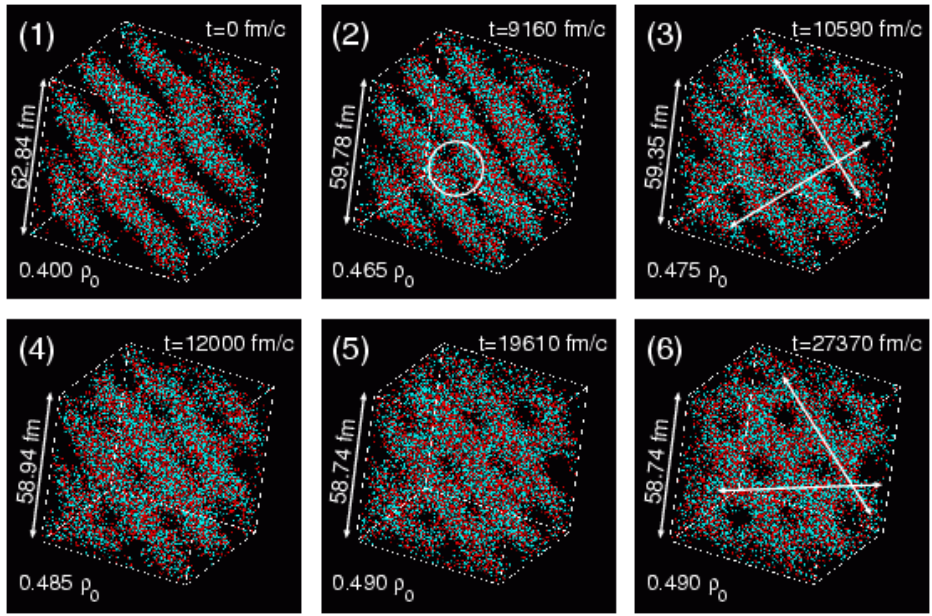}}
\caption{\label{fig slab cylindhole}
The same as Fig.\ \ref{fig rod slab} for the transition from
the phase with slab-like nuclei to the phase with cylindrical holes
(the box size is not rescaled in this figure).
Adapted from Ref.\ \cite{qmd_transition}.
}
\end{figure}

The resulting time evolution of the nucleon distribution
is shown in Figs.\ \ref{fig rod slab} and \ref{fig slab cylindhole}.
In Fig.\ \ref{fig rod slab},
we see that the phase with slab-like nuclei is finally formed 
[Fig.\ \ref{fig rod slab}-(6)]
from the phase with rod-like nuclei [Fig.\ \ref{fig rod slab}-(1)].
We note that the transition is triggered by thermal fluctuations,
not by the fission instability:
when the internuclear spacing becomes small enough 
and once some pair of neighboring rod-like nuclei 
touch due to the thermal fluctuations,
they fuse [Figs.\ \ref{fig rod slab}-(2) and \ref{fig rod slab}-(3)].
Then such connected pairs of rod-like nuclei further touch and fuse with 
neighboring nuclei in the same lattice plane like a chain reaction
[Fig.\ \ref{fig rod slab}-(4)]; the time scale of the each fusion process
is $O(10^2)$ fm$/c$.

The transition from the phase with slab-like nuclei
to the phase with cylindrical holes is shown in 
Fig.\ \ref{fig slab cylindhole}.
When the internuclear spacing decreases enough,
neighboring slab-like nuclei touch due to the thermal fluctuation
as in the above case.
Once nuclei begin to touch [Fig.\ \ref{fig slab cylindhole}-(2)],
bridges between the slabs are formed at many places
on a time scale of $O(10^2)$ fm$/c$.
Initially, the bridges cross the slabs
almost orthogonally [Fig.\ \ref{fig slab cylindhole}-(3)].
Nucleons in the slabs continuously flow into the bridges and the bridges
become wider and merge together to form cylindrical holes.
Then the cylindrical holes finally relax into a triangular lattice 
[Fig.\ \ref{fig slab cylindhole}-(6)].

\section{IV.\quad Summary and Future Prospects}

Using QMD, we have shown the following two things:
formation of the pasta phases in neutron star crusts by cooling and
formation of rod-like bubbles from slab-like nuclei and that of
slab-like nuclei from rod-like ones in supernova cores 
by compression of matter.
In closing, we list important issues to be clarified in the future.
\begin{itemize}
\item[1.] Formation of the pasta phases by compression \cite{future}\\
A remaining problem is a transition from a bcc lattice of 
spherical nuclei to a triangular lattice of rod-like nuclei 
induced by compression. If this process is
confirmed, existence of the pasta phases in supernova
cores will be almost established.

\item[2.] Effects of uncertainties of nuclear forces \& properties of nuclei\\
We employed a specific nuclear force.
Although we have confirmed that 
this nuclear interaction yields reasonable values for the surface energy and
the proton and neutron chemical potentials in neutron matter, etc. 
\cite{qmd_cold},
in addition to reproducing saturation properties \cite{maruyama},
further systematic survey is needed to understand effects of uncertainties
in nuclear properties on the pasta phases \cite{sonoda_future}.
Especially, uncertainties in the surface properties 
are the most important elements to be examined.

\item[3.] Detailed study of astrophysical consequences\\
If formation of the pasta phases in actual astrophysical situations is
established, quantitative discussion about effects of the pasta phases
on astrophysical phenomena will be increasingly important.
As we have stressed previously \cite{gentaro,qmd_cold},
effects of the pasta structure on coherent scattering of neutrinos in
the stage of the neutrino trapping is an interesting problem 
\cite{horowitz,qmd_hot,sonoda_future,sonoda}.

\end{itemize}


\begin{theacknowledgments}
The author is grateful to Chris Pethick for helpful comments.
The research reported in this article grew out of collaborations with
H. Sonoda, K. Sato, T. Maruyama, K. Yasuoka, T. Ebisuzaki, and K. Iida.
Further research currently in progress is performed
using the RIKEN Super Combined Cluster System with MDGRAPE-2.
This work was supported in part
by a JSPS Postdoctoral Fellowship for Research Abroad,
by the Nishina Memorial Foundation,
by the Japan Society for the Promotion of Science,
by the Ministry of
Education, Culture, Sports, Science and Technology
through Research Grant No. 14-7939,
and by RIKEN through Research Grant No. J130026.
\end{theacknowledgments}


\begin{thebibliography}{99}

\bibitem{review} C. J. Pethick and D. G. Ravenhall,
  Annu.\ Rev.\ Nucl.\ Part.\ Sci. {\bf 45}, 429 (1995).
%
\bibitem{chamel} N. Chamel,
  contribution to this volume.
%
\bibitem{rpw} D. G. Ravenhall, C. J. Pethick, and J. R. Wilson,
  Phys.\ Rev.\ Lett. {\bf 50}, 2066 (1983).
%
\bibitem{hashimoto} M. Hashimoto, H. Seki, and M. Yamada,
  Prog.\ Theor.\ Phys. {\bf 71}, 320 (1984).
%
\bibitem{williams} R. D. Williams and S. E. Koonin,
  Nucl.\ Phys. {\bf A435}, 844 (1985).
%
\bibitem{lassaut} M. Lassaut {\it et al.},
  Astron.\ Astrophys. {\bf 183}, L3 (1987).
%
\bibitem{lorenz} C. P. Lorenz, D. G. Ravenhall, and C. J. Pethick,
  Phys.\ Rev.\ Lett. {\bf 70}, 379 (1993).
%
\bibitem{oyamatsu} K. Oyamatsu,
  Nucl.\ Phys. {\bf A561}, 431 (1993).
%
\bibitem{sumiyoshi} K. Sumiyoshi, K. Oyamatsu, and H. Toki,
  Nucl.\ Phys. {\bf A595}, 327 (1995).
%
\bibitem{gentaro} G. Watanabe, K. Iida, and K. Sato,
  Nucl.\ Phys. {\bf A676}, 455 (2000);
  {\it ibid.} {\bf A687}, 512 (2001);
  Erratum, {\it ibid.} {\bf A726}, 357 (2003).
%
\bibitem{aichelin} J. Aichelin and H. St{\"o}cker,
  Phys.\ Lett. {\bf B176}, 14 (1986); J. Aichelin,
  Phys.\ Rep. {\bf 202}, 233 (1991).
%
\bibitem{qmd_cold} G. Watanabe {\it et al.},
  Phys.\ Rev.\ C {\bf 66}, 012801(R) (2002); 
  {\it ibid.} {\bf 68}, 035806 (2003).
%
\bibitem{qmd_hot} G. Watanabe {\it et al.},
  Phys.\ Rev.\ C {\bf 69}, 055805 (2004).
%
\bibitem{qmd_transition} G. Watanabe {\it et al.},
  Phys.\ Rev.\ Lett. {\bf 94}, 031101 (2005).
%
\bibitem{shell p} K. Oyamatsu and M. Yamada,
  Nucl.\ Phys. {\bf A578} (1994) 181.
%
\bibitem{shell n} P. Magierski and P. -H. Heenen,
  Phys.\ Rev.\ C {\bf 65} (2002) 045804.
%
\bibitem{chamel_shell} N. Chamel,
  Nucl.\ Phys. {\bf A747}, 109 (2005).
%
\bibitem{maruyama} T. Maruyama {\it et al.},
  Phys.\ Rev.\ C {\bf 57}, 655 (1998).
%
\bibitem{kido} T. Kido {\it et al.},
  Nucl.\ Phys. {\bf A663 \& 664}, 877c (2000).
%
\bibitem{niita} K. Niita,
  in the Proceedings of the Third Simposium on
  {\it ``Simulation of Hadronic Many-body System''},
  A. Iwamoto {\it et al.}, Eds.,
  JAERI-conf. {\bf 96-009}, 22 (1996) (in Japanese).
%
\bibitem{screening} G. Watanabe and K. Iida,
  Phys.\ Rev.\ C {\bf 68}, 045801 (2003).
%
\bibitem{maruyama_screen} T. Maruyama {\it et al.},
  Phys.\ Rev.\ C {\bf 72}, 015802 (2005).
%
\bibitem{bbp} G. Baym, H. A. Bethe, and C. J. Pethick,
  Nucl.\ Phys. {\bf A175}, 225 (1971).
%
\bibitem{horowitz} C. J. Horowitz {\it et al.},
  Phys.\ Rev.\ C {\bf 70}, 065806 (2004).
%
\bibitem{soft_review} G. Watanabe and H. Sonoda,
  to appear in ``Soft Condensed Matter: New Research'',
  ed. K. I. Dillon (cond-mat/0502515).
%
\bibitem{matsuzaki} M. Matsuzaki,
  Phys.\ Rev.\ C {\bf 73}, 028801 (2006).
%
\bibitem{future} G. Watanabe {\it et al.},
  in preparation.
%
\bibitem{sonoda_future} H. Sonoda {\it et al.},
  in preparation.
%
\bibitem{sonoda} H. Sonoda,
  contribution to this volume.
%
\end{thebibliography}
\end{document}